\documentclass[9pt,twocolumn,twoside]{osajnl}

\journal{ol} 

\setboolean{shortarticle}{true}

\usepackage{siunitx}
\usepackage[normalem]{ulem}

\DeclareMathOperator{\var}{\text{var}}
\DeclareMathOperator{\std}{\text{std}}

\title{Compact and high-precision wavemeters using the Talbot effect and signal processing}

\author[*]{Ningren Han}
\author[ \hspace{-0.8ex}]{Gavin N. West}
\author[ \hspace{-0.8ex}]{Amir H. Atabaki}
\author[ \hspace{-0.8ex}]{David Burghoff}
\author[ \hspace{-0.8ex}]{Rajeev J. Ram}

\affil[ \hspace{-0.8ex}]{Department of Electrical Engineering and Computer Science, Massachusetts Institute of Technology, Cambridge, Massachusetts 02139, USA}

\affil[*]{Corresponding author: hanningren@alum.mit.edu}

\begin{abstract}

Precise knowledge of a laser's wavelength is crucial for applications from spectroscopy to telecommunications. Here we present a wavemeter which operates on the Talbot effect. Tone parameter extraction algorithms are used to retrieve the frequency of the periodic signal obtained in the sampled Talbot interferogram. Theoretical performance analysis based on the Cram\'er-Rao lower bound (CRLB) as well as experimental results are presented and discussed. With this scheme, we experimentally demonstrate a compact and high-precision wavemeter with below $10$ pm single-shot estimation uncertainty under the 3-$\sigma$ criterion around $780$ nm.

\end{abstract}

\setboolean{displaycopyright}{true}

\begin{document}

\maketitle

The Talbot effect refers to the self-imaging phenomenon observed with coherent light after passing through periodic structures such as diffraction gratings \cite{talbot1836lxxvi}. Owing to its interferometric nature with an invertible spatio-spectral response, it has been previously proposed as the building component for realizing spectrometers \cite{kung2001transform,de2004talbot}. With modern technological advances in CCD and CMOS image sensors, sensor pixel sizes have reached to a point where direct sampling of the Talbot pattern is achievable without any external imaging optics \cite{ye2016miniature,han2016non}. In addition, operating the spectrometer configuration in the non-paraxial regime with a tilted image sensor allows full-frame interferogram capture without any moving parts, as strong diffractions in the non-paraxial regime confine the Talbot region to be within several millimeters after the diffraction grating. This creates a simple and compact spectrometer configuration, while still maintaining a competitive performance in terms of spectral resolution, bandwidth, and throughput \cite{ye2016miniature}.

Aside from spectrometers designed for general spectroscopy (such as those designed for broadband light sources), more specialized wavelength meters play an important role in applications such as spectroscopy of atomic systems. Also known as wavemeters, they are specifically designed for measuring the wavelength of coherent laser beams. A wavemeter usually has resolution requirements much higher than that of a typical spectrometer and is generally built through an interferometric geometry. One of the key factors to a good wavemeter performance is to ensure low estimation uncertainty across measurements. This is referred to as high precision in our text. The commercial landscape for wavemeters has been dominated by two approaches: the scanning Michelson interferometer and the static Fizeau interferometer. The highest-precision wavemeters are typically based on the Fizeau geometry \cite{gray1986simple}. In addition, it has better performance against power fluctuations and side modes due to the static nature. Michelson interferometer-based wavemeters, on the other hand, can cover longer wavelength regimes due to the fact that only a single detection element is required \cite{fox1999reliable}. Apart from these conventional approaches, in recent years, new concepts for realizing computational devices and instruments have also resulted in a number of compact, high-precision, and broad-bandwidth wavemeters \cite{mazilu2014random,metzger2017harnessing,cao2017perspective}.

\begin{figure}[!htb]
  \centering
  \includegraphics[width=\linewidth]{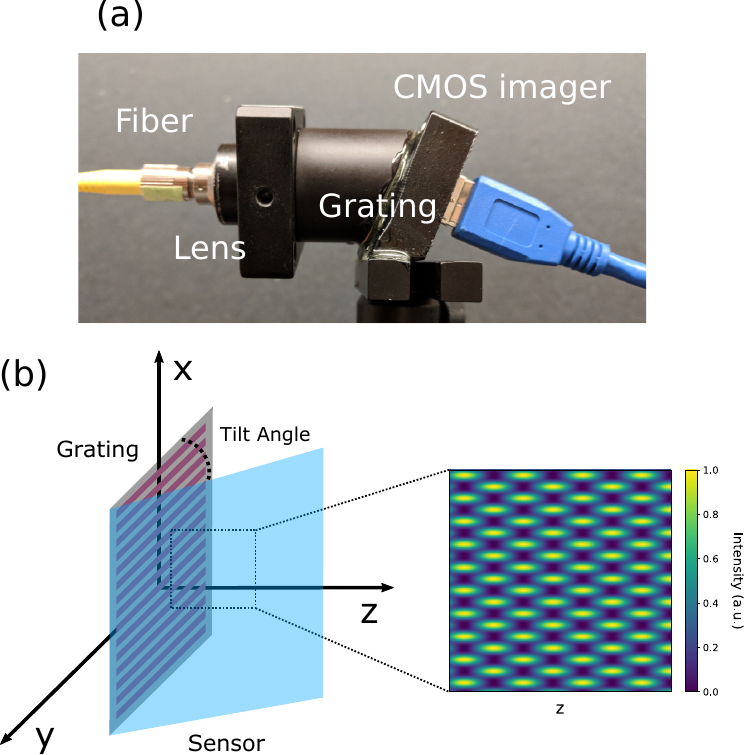}
  \caption{(a) The compact Talbot wavemeter. (b) Illustration of Talbot interferogram sampling with a tilted image sensor in close proximity to the grating.}
  \label{fig1}
\end{figure}

As an extension to our previous work on compact Talbot spectrometers \cite{ye2016miniature,han2016non}, here we utilize the Talbot effect and signal processing for realizing wavemeters, as shown in Fig.~\ref{fig1}~(a). With $P$ as the grating period and $\lambda$ as the incidence light wavelength, the Talbot distance $z_T$, over which the self images repeat, is given as
\begin{equation}
  z_T = \frac{\lambda}{1 - \sqrt{1 - \left( \frac{\lambda} {P} \right)^2}}.
  \label{talbot_distance_full}
\end{equation}
\noindent This was first discovered in 1836 by Henry Fox Talbot \cite{talbot1836lxxvi}, and later analyzed by Lord Rayleigh \cite{rayleigh1881xxv}. The Talbot periodicity in the depth ($z$) direction provides a spatio-spectral mapping that is interferometric in nature. As shown in Fig.~\ref{fig1}~(b), a tilted image sensor can be placed after the grating to sample the Talbot interferogram across depths. Afterwards, a straightforward Fourier transform can be used to reconstruct the spectrum of the light source \cite{kung2001transform,de2004talbot,ye2016miniature}.

For coherent light signals such as laser sources, the sampled Talbot interferogram rows (across the depth dimension) essentially contain periodic signals where the spatial periodicity corresponds to the laser wavelength according to Eq.~\ref{talbot_distance_full}. Extracting the frequency (and possibly amplitude and phase) information for a periodic signal, which some term as the tone parameter estimation problem \cite{rife1974single}, can be achieved with precisions much higher than those obtained from direct FFT of the spectrogram \cite{rife1974single}. This problem has a long history in the signal processing community with applications ranging from radar and sonar systems \cite{schmidt1986multiple,roy1989esprit}, audio and acoustics \cite{smith1987parshl,de2002yin}, astronomy \cite{gregory2005bayesian}, and many more. As a result, many algorithms such as the maximum likelihood estimation \cite{rife1974single}, MUSIC (MUltiple SIgnal Classification) \cite{schmidt1981signal,schmidt1986multiple}, ESPRIT (Estimation of Signal Parameters via Rotational Invariance Technique) \cite{roy1989esprit}, and others \cite{boashash1992estimating} have been developed and applied to solve this problem with extremely high precision.

\begin{figure}[!htb]
  \centering
  \includegraphics[width=\linewidth]{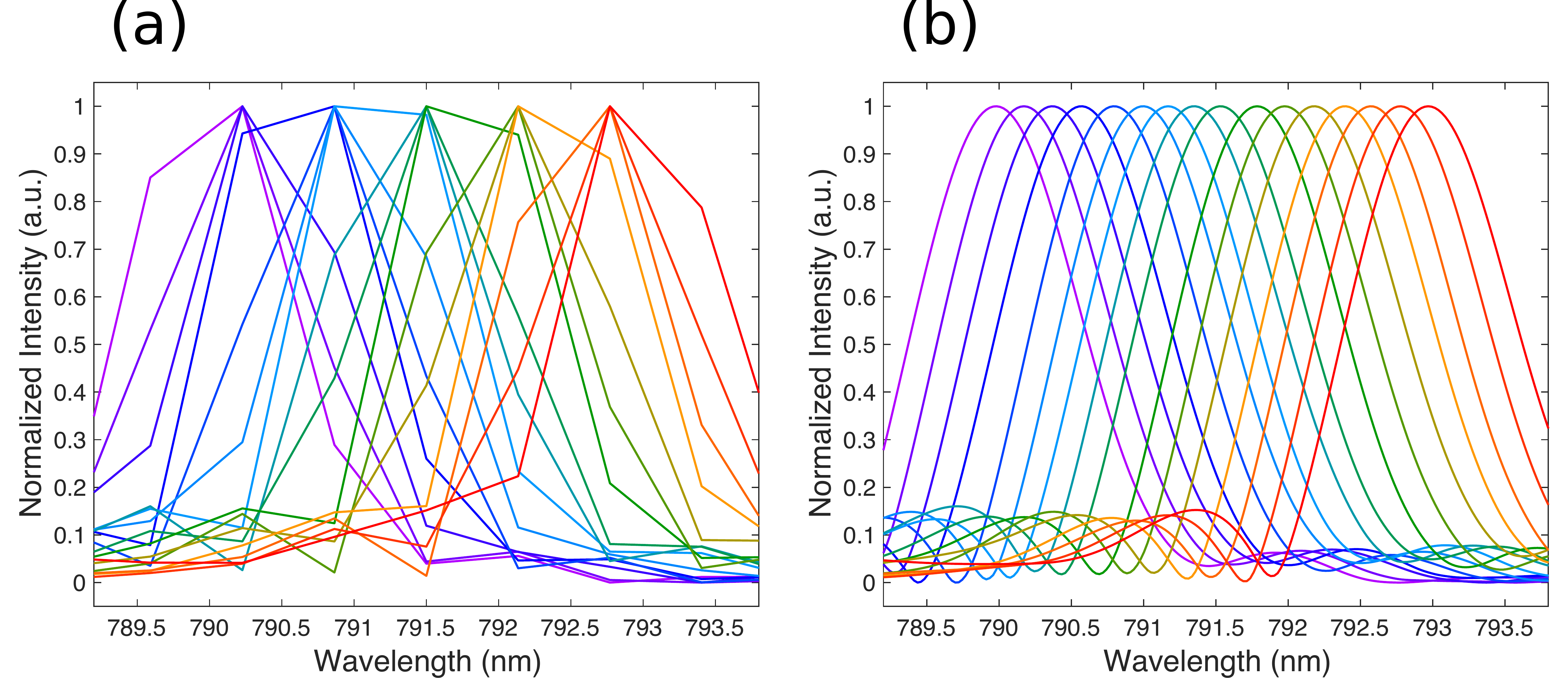}
  \caption{(a) Reconstructed spectra using direct FFT for stepping laser wavelengths with a tunable external cavity diode laser. (b) Reconstructed spectra using FFT with prior zero-padding for the spectral sources in (a).}
  \label{fig2}
\end{figure}

While direct FFT has been used for spectrum retrieval with the Talbot spectrometer in our previous work, for laser wavelength estimation, similar tone parameter extraction ideas can be applied to achieve much higher laser wavelength estimation precisions. This is illustrated in Fig.~\ref{fig2}, which shows example plots for reconstructed spectra for stepping laser wavelengths across a small wavelength range. Fig.~\ref{fig2}~(a) uses direct FFT for spectral processing, whereas Fig.~\ref{fig2}~(b) zero-pads the interferogram rows to augment the array sizes by around four times prior to the FFT operation for precision enhancement. As can be seen from the figure, with zero-padding, much finer interpolated spectral shapes can be achieved, resulting in much more precise center frequency estimation than that from the direct FFT estimation.

For real sinusoidal parameter estimation, assume that the underlying periodic signal $y[n]$ is
\begin{equation}
  y[n] = A \cos (2 \pi f_0 \Delta n + \phi),
\end{equation}
for $n = 0, 1, 2, ..., N - 1$, where $A$ is the amplitude for the periodic signal, $f_0$ is its frequency, $\Delta$ is the sampling interval, and $\phi$ is the phase. The observed discrete noisy signal $x[n]$ is
\begin{equation}
  x[n] = y[n] + w[n],
\end{equation}
\noindent where $w[n]$ is white Gaussian noise with variance $\sigma^2$. The Cram\'er-Rao lower bound (CRLB), which is the theoretical lowest error bound achievable with an unbiased estimator, for the various parameters in the model is \cite{sengijpta1995fundamentals}
\begin{equation}
  \var [\hat {A}] \ge \frac{2 \sigma^2} {N},
\end{equation}
\begin{equation}
  \var [\hat {f_0}] \ge \frac {6 \sigma^2} {\pi^2 A^2 \Delta^2 N (N^2 - 1)},
\end{equation}
\begin{equation}
  \var [\hat {\phi}] \ge \frac{4 (2 N - 1) \sigma^2} {A^2 N (N + 1)}.
\end{equation}
\noindent Here, $\hat{\cdot}$ refers to the estimator for the corresponding parameter. The most relevant parameter for our application is $\var [\hat {f_0}]$. The model SNR, which is usually defined as the ratio between the variance of the signal and the variance of the noise, is
\begin{equation}
  \text{SNR} = \frac{\var[y]}{\var[w]} = \frac{A^2} {2 \sigma^2}.
\end{equation}
\noindent Rewriting the CRLB for frequency estimation, we have
\begin{equation}
  \var [\hat {f_0}] \ge
  \frac {3} {\pi^2 \, \text{SNR} \, \Delta^2 N (N^2 - 1)} =
  \frac {3 N (\delta f)^2} {\pi^2 \, \text{SNR} \, (N^2 - 1)}.
\end{equation}
\noindent Here, $\delta f$ is the FFT-defined frequency domain spacing. The standard deviation for the frequency estimation error can therefore be written as
\begin{equation}
  \std [\hat {f_0}] \gtrapprox \frac {\delta f} {\sqrt{3 \, \text{SNR} \, N}}.
  \label{frequency_estimation_std}
\end{equation}

Eq.~\ref{frequency_estimation_std} can be used to estimate the theoretical wavelength precision bound for a Talbot wavemeter. Taking the Aptina MT9J003 image sensor as an example, which has dimensions of $3856 \times 2764$, a moderate experimental SNR of $\approx 0.5$ for a single-row interferogram signal (across the larger dimension) is generally achievable in our experiments. This corresponds to a 3-$\sigma$ estimation precision of $\approx 4 \%$ of the FFT bin size $\delta f$. If ensemble estimation such as mean aggregation based on the row-wise estimations is performed, a further reduction of $\sqrt{M}$ times in estimation uncertainty can be achieved for perfect and uniform plane wave incidence sources, where $M$ is the number of rows. With $\delta f$ below $1$ nm for most geometries tested under optical wavelengths in our configurations \cite{ye2016miniature}, this means that single-shot sub-picometer precision based on the 3-$\sigma$ criterion is possible with image sensors like Aptina MT9J003. In practice, issues such as wavefront aberration and non-uniform incidence power spread can introduce additional estimation biases and uncertainties, thereby deviating the actual system performance from the theoretical estimation bound. However, the analysis here can still provide insights on the system performance limits and help identify sources for improvements.

Experiments to explore the performance limits for the Talbot wavemeter and signal processing for wavelength estimation were carried out. A compact Talbot wavemeter setup using 3-D printed parts and Thorlabs components similar to the one shown in Fig.~\ref{fig1}~(a) was used for device characterization. A tunable single-frequency Ti:Sapphire laser (SolsTiS, M-Squared Lasers) was used as the light source. A high-resolution wavemeter (HighFinesse WS7) was used to provide the reference wavelength measurements with sub-picometer accuracies. A single-mode optical fiber was used for light delivery. A fiber collimation lens and a beam expander were used to fill the image sensor area. The Talbot system consisted of the Aptina MT9J003 sensor and a diffraction grating with $1.035$ \textmu m grating pitch size. The grating-sensor tilt angle was $\ang{20}$. The laser source was tuned across pre-defined wavelength ranges. For each wavelength, consecutive and independent images were obtained for estimation variance analysis.

\begin{figure}[!htb]
  \centering
  \includegraphics[width=\linewidth]{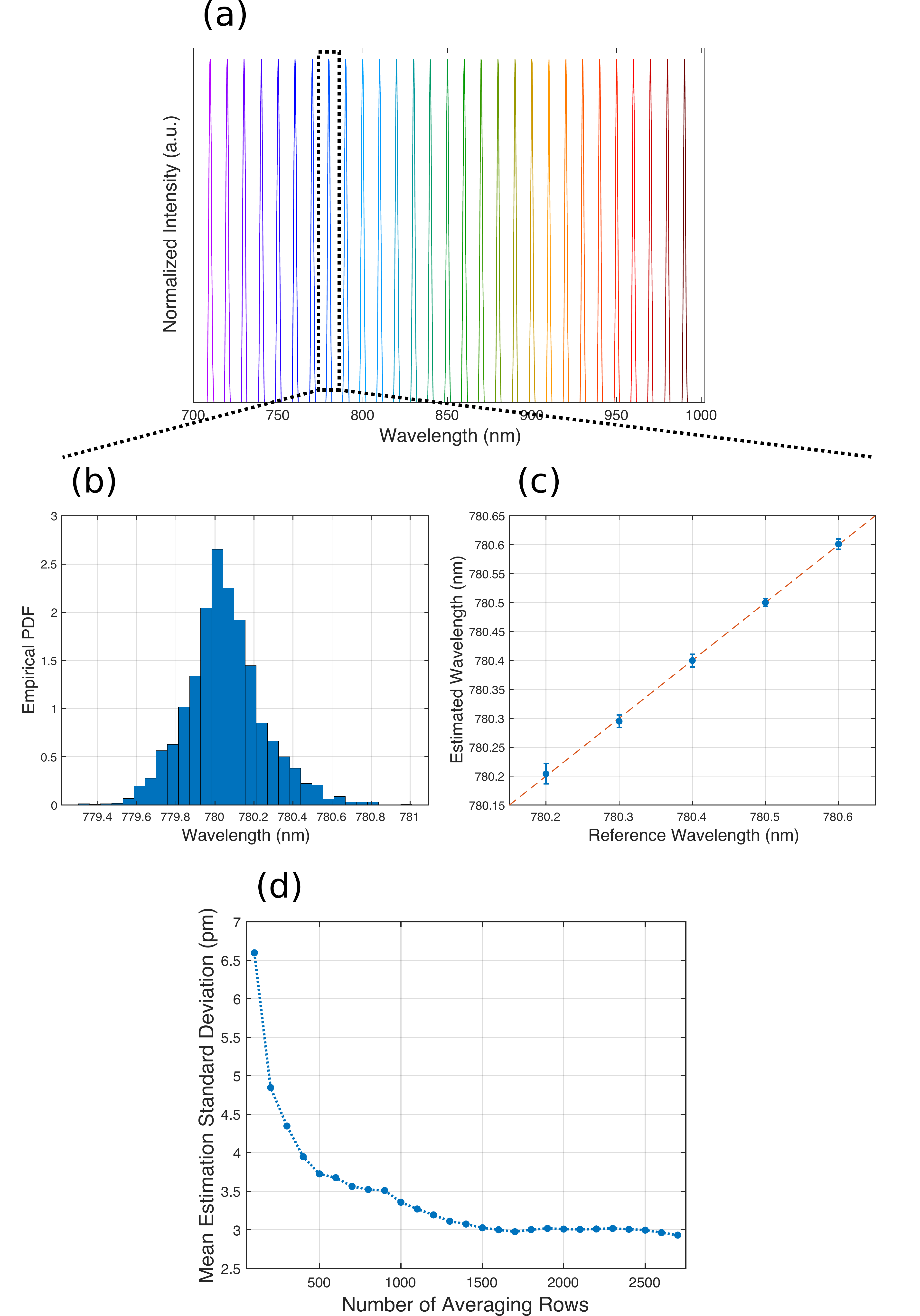}
  \caption{(a) FFT-reconstructed spectral peaks for a single-frequency Ti:Sapphire source tuned from $710$ nm to $990$ nm in steps of $10$ nm. (b) The empirical PDF for row-wise wavelength estimations across the captured image at $780$ nm. (c) Wavelength estimations for a laser source tuned with $100$ pm step sizes. The algorithm used here was the FFT peak localization method. The dots in the plots are the means from $10$ consecutive and independent acquisitions. The error bars represent the 3-$\sigma$ uncertainties from the $10$ measurements. (d) Mean estimation standard deviation as a function of the number of averaging rows for the MUSIC algorithm. The standard deviations were calculated from $100$ acquisitions with $10$ different wavelength points around $780$ nm.}
  \label{fig3}
\end{figure}

We first show the broad bandwidth operation with the Talbot wavemeter in Fig.~\ref{fig3}~(a). The Ti:Sapphire source was tuned from $710$ nm to $990$ nm in steps of $10$ nm and the corresponding FFT-reconstructed spectral peaks are shown in Fig.~\ref{fig3}~(a). In theory, the Talbot wavemeter is able to cover the entire optical wavelength range as long as the sensor pixel remains responsive. In practice, our experimental demonstration is limited by the source availability. Due to the finite experimental control precision for the light incidence angle and the grating-sensor tilt angle, a direct wavelength estimation based on the sampled interferogram according to Eq.~\ref{talbot_distance_full} does not match exactly with the actual wavelength. As a result, a linear calibration based on the least squares fitting between the calculated wavelengths and the actual wavelengths was carried out similar to \cite{ye2016miniature}.

To estimate the wavelength from the Talbot interferogram image with high precision, row-wise wavelength estimation was carried out for all image rows across which depth samplings were performed. Two algorithms have been extensively explored. The first one was an algorithm based on peak localization with zero-padded FFT \cite{smith1987parshl}. The interferogram rows were first zero-padded to augment the array dimension by one to two orders of magnitude. FFT was then applied on the signal for spectrum retrieval. Afterwards, the maximum of the spectral peak was identified, and a parabolic approximation based on this point and its adjacent points was used for peak maximum localization. The second one was the MUSIC algorithm \cite{schmidt1981signal,schmidt1986multiple}, which is an eigenspace method to identify a known number of sinusoidal signals in the presence of Gaussian white noise. It is considered by many as one of the most promising algorithms for frequency estimation tasks \cite{barabell1998performance}. For this algorithm, we used the MATLAB implementation (rootmusic) for our numerical processing. A bandpass filter was applied prior to the MUSIC algorithm to filter out unwanted signal interferences such as the reflection from the glass window over the image sensor. Fig.~\ref{fig3}~(b) shows the empirical probability density function (PDF) for row-wise wavelength estimations across the captured image with the MUSIC algorithm at $780$ nm. As mentioned in the previous text, mean aggregation was then applied to reduce the estimation uncertainty.

With the above approaches, both algorithms were able to provide accurate wavelength estimations much better than those obtained through direct FFT inversion. To quantify the estimation uncertainty, which defines the precision achievable with our approach, we used $10$ different wavelength measurements around $780$ nm, each with $10$ consecutive and independent acquisitions, to calculate the mean estimator standard deviation. As shown in Tab.~\ref{tab1}, the final single-shot mean 3-$\sigma$ uncertainty was $\approx 9.8$ pm for the FFT peak localization algorithm and $\approx 8.8$ pm for the MUSIC algorithm. Both algorithms were able to provide sub-10 picometer single-shot estimation precision based on the 3-$\sigma$ criterion, with the MUSIC algorithm having a slightly more consistent estimation. Fig.~\ref{fig3}~(c) shows wavelength estimations with the FFT peak localization algorithm for steps of $100$ pm with 3-$\sigma$ as the error bar. A linear wavelength calibration similar to the one applied in Fig.~\ref{fig3}~(a) was used for this wavelength range prior to the plot.

\begin{table}[!htb]
  \centering
  \scalebox{0.85} {
    \begin{tabular}{c c c c c c c c c c c c}
      \toprule
      Algorithm & FFT peak localization & MUSIC \\
      \midrule
      3-$\sigma$ uncertainty & 9.8 pm & 8.8 pm \\
      \bottomrule
    \end{tabular}
  }
  \caption{Single-shot 3-$\sigma$ uncertainties for the FFT peak localization algorithm and the MUSIC algorithm.}
  \label{tab1}
\end{table}

The effect of mean aggregation across different interferogram rows is investigated next. This is shown in Fig.~\ref{fig3}~(d), where we varied the number of averaging rows from $100$ to $2700$ in steps of $100$ with the MUSIC algorithm. As can be seen from the plot, mean aggregation across the interferogram rows can enhance the estimation precision considerably. The most significant improvement comes from the initial aggregation stage, which is expected for uncertainty reduction with averaging. The CRLB suggests a sub-picometer single-shot estimation standard deviation with full mean aggregation. The performance gap from theory is likely caused by the fact that in our experiment, aberrations with the collimation setup as well as the non-ideal pixel sampling either due to the oblique incidence or the micro-lens array can introduce phase errors in our interferogram signal. This causes Fourier-domain spectral peak distortions as well as peak position misalignments across different interferogram rows, weakening the efficacy of mean aggregation in terms of uncertainty reduction. This can likely be improved by better collimation for aberration reduction or by using algorithms that can correct systematic phase errors during estimation.

While the MUSIC algorithm yielded better precisions in terms of estimation consistency, a significant advantage for the FFT peak localization algorithm is its computational speed. While extra memory is needed for storing the zero-padded image, row-wise FFT across a two-dimensional image has vectorized and well-optimized code executions \cite{frigo2005design}. The remaining operations generally have linear time complexities and are easily vectorized. As discussed earlier, aggregating row-wise estimation results is one of the keys to achieve an accurate final estimation, therefore being able to perform fast and parallel frequency estimations across several thousand interferogram rows can be extremely advantageous. In general, our experiments suggested at least $\approx 20$ times faster computational speed for the FFT localization algorithm as compared to the MATLAB-provided MUSIC algorithm. In addition, the MUSIC algorithm requires the number of existing periodic signals to be known in advance, which would need more care when multiple sources exit. This aspect can be handled more easily with the FFT localization algorithm, as one can incorporate a heuristic approach for peak identification and thresholding within the FFT localization steps in a straightforward manner.

The approach presented in this work provides an attractive option for realizing compact and high-precision wavemeters. While recent computational wavemeter approaches have demonstrated impressive size, resolution, and bandwidth metrics \cite{mazilu2014random,metzger2017harnessing}, these approaches generally require a full-spectrum calibration process to explicitly construct the system transfer matrix. This may be difficult to have in practical use cases. On the other hand, the interferometric nature of the Talbot phenomenon presents recorded signals in analytically-tractable periodic forms. This can potentially eliminate the need for a full-spectrum calibration and can leverage many canonical signal processing algorithms for speedy and accurate parameter estimation. Compared to the Fizeau or Michelson interferometer-based wavemeters, the static and more compact Talbot geometry proposed in our work is simple to realize with inexpensive optical components such as diffraction gratings and CMOS image sensors. As a result, it can be a preferred choice where portability, speed, and cost are the constraining factors in the application.

As a non-negligible performance gap exists in terms of the actual wavelength estimation precision and the theoretical bound, some improvements can be made. For example, one can perform better wavefront shaping and improve the power uniformity for the incident light. In addition, while a high absolute accuracy is shown in Fig~\ref{fig3}~(c) across a narrow wavelength range, a precise geometry and robust calibration scheme for the Talbot wavemeter has yet to be performed over a wide wavelength range. As an example, a quadratic wavelength calibration across a $100$ nm wavelength span around $780$ nm yielded an absolute wavelength accuracy error of around $40$ pm with the current configuration. While this is largely a solved technical issue given the industry success of interferometer-based wavemeters, improving the Talbot wavemeter for practical usage in this regard may still take several design iterations and adjustments. Last but not least, improving the sampled Talbot interferogram visibility will lead to directly increased SNR. This can be achieved through experimenting with new sensor-grating combinations.

\textbf{Funding}: This project is funded by the Disruptive \& Sustainable Technologies for Agricultural Precision (DiSTAP) program under the Singapore-MIT Alliance for Research and Technology (SMART).

\bibliography{paper}

\bibliographyfullrefs{paper}

\end{document}